# An Efficient Approach to Detecting Lung Nodules Using Swin Transformer

Saeed Shakuri
Department of Computer Engineering
University of Science and Culture
Tehran, Iran
saeed.shakuri@stu.usc.ac.ir

Alireza Rezvanian
Department of Computer Engineering
University of Science and Culture
Tehran, Iran
rezvanian@usc.ac.ir

*Abstract*— Lung cancer has the highest rate of cancer-caused deaths, and early-stage diagnosis could increase the survival rate. Lung nodules are one of the most common lung cancer signs, therefore, the development of lung nodule detection systems becomes significantly crucial. Different lung nodule detection models have been introduced, however, they haven't fully considered all aspects of efficiency. Hence, in this paper, we take a more efficient approach in introducing a lung nodule detection model while providing improvements in nodule detection. In this regard, we consider 2D slices of the CT scans which not only lead to lower computational load and complexity in training and inference phases but also, allow us to examine the potential of 2D models in lung nodule detection. Moreover, we utilize the tiny version of Swin Transformer to leverage the benefits of Vision Transformers (ViT) while having a much lower computational complexity. We also build a Feature Pyramid Network on top of our backbone to detect nodules of various sizes better, especially small nodules. Additionally, we consider Transfer Learning for the training procedure, resulting in significantly lower training iterations. The experimental results demonstrate the competitive performance of our model compared to the state-of-the-art models with even higher mAP and mAR for small nodules by 1.3% and 1.6% respectively. Also, our proposed model achieved the highest mAP in all size nodules with 94.7% and mAR of 94.9%.

## I. INTRODUCTION

The most common cause of cancer death is lung cancer [1]. According to the *National Cancer Center* data in 2022, the lung cancer death accounts for 20.4% of all cancer deaths. Due to the lack of early-stage diagnosis, the 5-year survival rate of lung cancer is less than 20% [2]. Hence, early-stage detection of lung cancer is significantly crucial for increasing survival rate [3]. One of the features of early-stage lung cancer is the appearance of lung nodules, which are lesions with a high probability of becoming malignant tumors [1]. If these nodules are not diagnosed promptly, they will turn into lumps which bring difficult and complicated treatment challenges [2]. Thoracic computed tomography (CT) is considered one of the best ways to detect nodules since it can provide high-quality slices of the lungs. However, nodules could appear in small sizes, making the detection challenging even for doctors to distinguish them from a great number of CT slices [1]. Therefore, automatic lung nodule detection could bring ease for radiologists in locating suspicious lung nodules [4]. Traditional lung nodule detection methods have gradually been replaced by deep learning-based approaches, which have better feature extraction capabilities and achieve better performance [5]. Object detection methods are usually divided into two categories, one-stage detections, which provide only one module for object localization and classification, and two-stage detections, which have two separate modules for localization and classification [6]. Two-stage object detection models have achieved high performance, and they also provide more flexibility in model architecture. Also, it has been reported that they provide better performance in small objects [7], which is crucial for lung nodule detection since they mostly appear as small lesions.

In CT scans, the scale of the Hounsfield Unit (HU) could be effective in preparing the CT scan slices to help the model focus more on the important objects according to the application. The different HU ranges for substances are shown in Table I. Also, CT scans provide a 3D structure of the lungs, and there have been many 3D models developed for lung nodule detection using these scans. However, 3D-based models require extremely high computational power for both training and inference stages and bring long training and detection times, which restricts their applications [8]. On the other hand, 2D models require much less computational power and are a more efficient choice for lung nodule detection, which also could bring more scalability in on-device implementations [8]. Furthermore, the potential of 2D models hasn't fully been examined as much as 3D models in lung nodule detection. However, there are different versions of 2D models, some of which could attain better performance in exchange for longer training and inference time as well as heavier models with more parameters and computational complexity. Hence, in an efficient approach, the choice among the model versions is a vital factor. Moreover, the training strategy could highly affect the efficiency of a model in terms of performance, inference phase, and training load, which is why it should be taken into consideration more carefully.

In this paper, we considered a more efficient approach to detecting lung nodules from CT scans. We introduce a 2D-based model using 2D axial slices of the LUNA16 dataset CT scans with Swin Transformer [9] and Feature Pyramid Network (FPN) [10] as the feature extractor instead of the conventional Convolutional Neural Networks (CNN). This model has the benefits of the Vision Transformers (ViT) [11] while having a lower computational complexity. Also, nodule features are extracted from feature maps of different layers of the backbone to provide a better understanding of the nodules of different sizes, especially small ones. Furthermore, Transfer Learning is used for the training stage, which would be a more efficient approach to training a model. Also, we examine the potential of 2D-based models in this application.

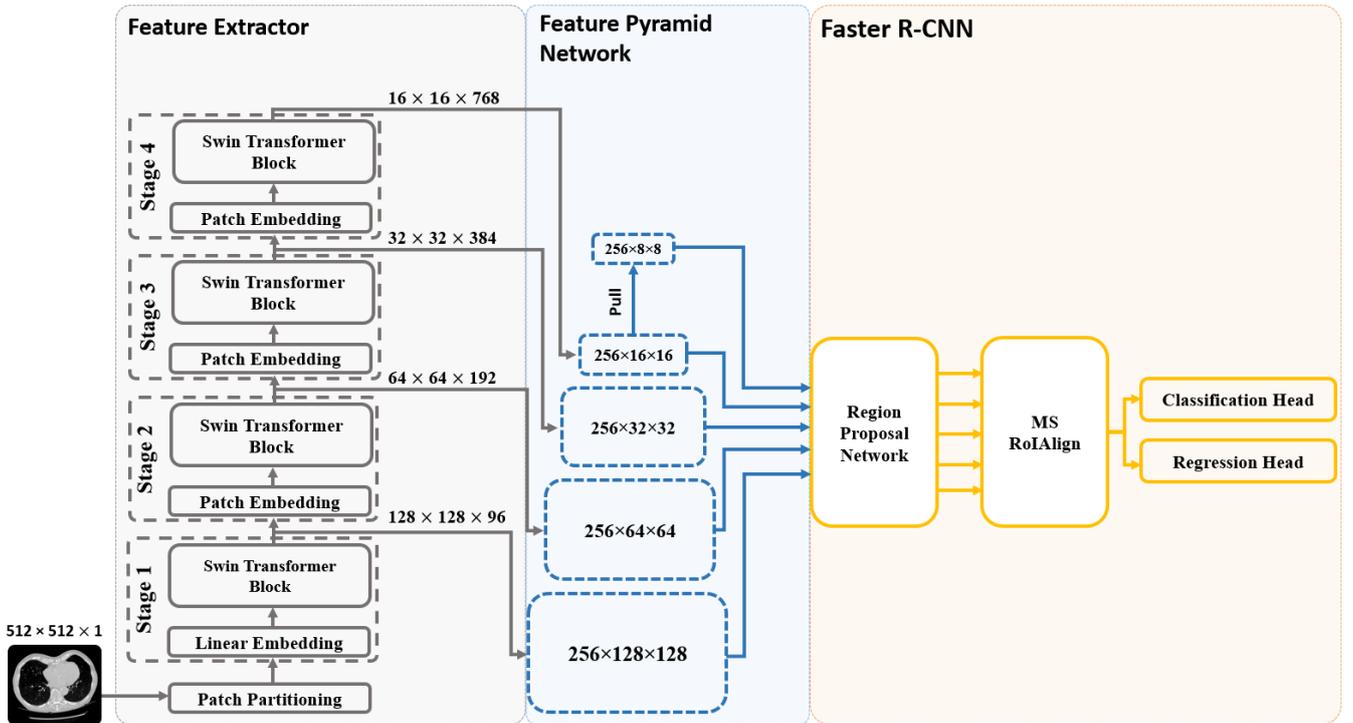

Fig 1. The overall structure of the proposed model. The grey area contains the Swin Transformer block followed by the blue area in which the FPN block is illustrated, and the orange area for the Faster RCNN stage.

The overall structure of the proposed model is illustrated in Fig. 1, and in-depth explanations are provided in the methodology in section III.

Additionally, we use the tiny version of the Swin Transformer as the backbone of the model, which has fewer parameters and lower computational complexity compared to the other Swin Transformer models. Also, this backbone has achieved better performance than most models with Convolutional Neural Network architecture [4]. The differences of the Swin Transformer models are in Table II.

## II. RELATED WORKS

In this section, the previous works considered different tracks in terms of model evaluation and methodologies, such as false positive reduction and one-stage detection, however, they used the LUNA16 dataset in train and test stages. The method proposed in [1] has a Slice Grouped Non-Local module to identify correlations in all channels and positions within slice groups while utilizing a 3D region proposal network (RPN) [13], and to compensate for the high false positive rate, a module is introduced to take advantage of the multi-scale feature map. The model was then evaluated with both 2D and 3D works under the false positive reduction track. The proposed model by Tong et al. [2] is also a CNN-based model that uses the Faster Region-based Convolutional Neural Network (Faster R-CNN) [14] method for object detection, and for candidate detection, an Iterative Self-Organizing Data Analysis Techniques Algorithm (ISODATA) is introduced which is a subsequent iteration version of $K$-means clustering algorithm and can split and merge the clustering results via the input distribution. Also, this model uses a Focal Loss to reduce the class imbalance of the dataset and evaluates the model in the false positive reduction track. Cui et al. [4] present a two-stage framework that uses a 3D Swin Transformer and FPN for the candidate nodule generation stage and a two-stream network to handle image and morphological features independently, which would remove false positives with high confidence. Moreover, an attention-based feature fusion module is used to combine the features of the LUNA16 dataset CT slices. In lung nodule detection previous papers, much less attention was paid to the potential of detection on 2D slices of the CT scans, among which Wu et al [8] considered this by introducing a multi-scale receptive module to overcome the blurred nature and irregular shape of the nodules in scans and using an Omni-dimensional convolution (EODConv) to improve the model ability in extracting filters and channels of the kernel while using an improved YOLOv7.

TABLE I. DIFFERENT SUBSTANCES' HU RANGES [12]

| Air | Lung | Fat | Water | Blood | Soft Tissue | Bone |
|---|---|---|---|---|---|---|
| -1000 | -500 | -100 to -50 | 0 | 30 to 45 | 100 to 300 | 700 to 3000 |

TABLE II. COMPARISON OF DIFFERENT SWIN TRANSFORMERS ON COCO CASCADE MASK R-CNN. "C" DENOTES FEATURE MAP CHANNEL SIZE, AND * INDICATES TRAINING WITH HTC++ [9]

| Models | Image size | C | Parameters | FLOPs | FPS |
|---|---|---|---|---|---|
| Swin-T | 224 | 96 | 86M | 756G | 15.3 |
| Swin-S | 224 | 96 | 107M | 838G | 12.0 |
| Swin-B | 224 | 128 | 145M | 982G | 11.6 |
| Swin-L* | 384 | 192 | 284M | 1470G | - |

As mentioned, the related works considered different aspects of lung nodule detection but did not fully consider all aspects of the efficiency of their models and the detection of small-size nodules which is the most common size. The nodule area distribution is shown in Fig. 2. Hence, not only improving the lung nodule detection performance is crucial but also, but it's important to have a more efficient approach in different aspects of a model, which could increase the scalability and practicality of the proposed model. Thus, the contribution of this paper is as follows:

- Using the 2D slices of 3D CT scans substantially reduces the computational costs in both training and inference.
- Employing the Swin-T to enhance the model performance while not increasing the computational complexity.
- Applying FPN to make use of feature maps of different layers of the backbone to detect all-size nodules better, especially small nodules.
- Utilizing Transfer Learning to train the model, which significantly reduces the training load.

### III. METHODOLOGY

The proposed model consists of a few steps, which will be discussed in more detail in this section.

*A. Data Preprocessing*

The data preprocessing stage has a few steps to utilize the 2D slices of the CT scans from their 3D structure, which are as follows:

- **Step 1:** Reading the scans, which are raw format files, and their annotations, followed by extracting certain information regarding the scans e.g. origin, spacing, and transform matrix for further process.
- **Step 2:** Walking through the axial slices of the scans, and extracting certain slices that contain nodules.
- **Step 3:** Since there are physical coordinates, conversion to voxel coordinates is crucial, which is carried out using origin, spacing, and transform matrix.
- **Step 4:** Finding corresponding annotations for the slices that contain nodules.
- **Step 5:** Performing windowing to rescale the pixel values to the Hounsfield Unit of -1000 to +400 [15] to fade the non-essential elements within the images, followed by saving the images as 12-bit grayscale pixels to cover the chosen HU range. Also, the converted annotations are saved in COCO format.

*B. Backbone*

Swin Transformer tiny is a type of Transformer-based model for computer vision tasks. In opposition to Transformers that work on the sequences, it processes the input through hierarchical feature maps. This hierarchical approach allows it to not only capture both fine and coarse-grained details efficiently but also, keep a linear computational complexity in contrast with ViT models which have quadratic computational complexity with respect to the input dimension [9]. In this research, the tiny version is used, which allows the model to balance the efficiency and performance with fewer parameters and model size.

The structure of the backbone consists of a few blocks. The Patch Partitioning block partitions the input image size of 512×512 by 4×4 patches, resulting in 128×128 patches, followed by the Linear Embedding block in which patches are flattened into an arbitrary dimension denoted as C=96, which leads to a dimension of 128×128×96. At this point, the input reaches the Swin Transformer block which applies a series of self-attention mechanisms and shifted window to capture both local and global information. The output of the first Swin Transformer block is passed to the next stage which starts with a Patch Merging block in which the input (128×128×96) is merged with 2×2 neighboring patches, and a linear layer is applied to make the feature dimension to 2C size, resulting in larger feature dimension and smaller spatial dimension (64×64×192). The output of this block is passed to a Swin Transformer block, and the trend continues from this point on, making four stages in total, and the final feature map with a size of 16×16×768.

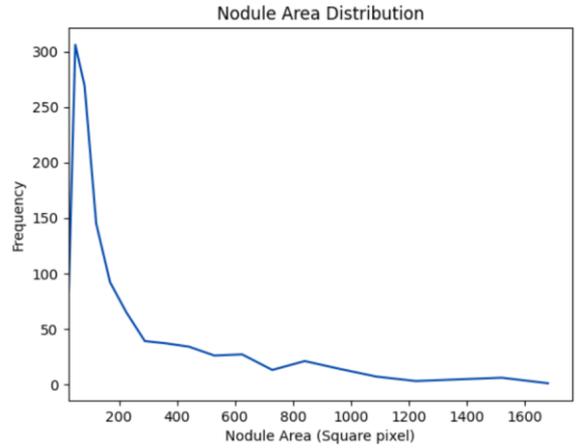

Fig 2. The distribution of the nodule areas.

After the backbone produces the feature maps, the FPN network exports feature maps from different layers of the backbone with a spatial dimension of 256. This way features from different layers of the feature maps are extracted, which is beneficial for detecting nodules of various sizes, especially small nodules. The feature maps are passed to the Faster R-CNN block in which the object detection phase takes place.

*C. Faster R-CNN*

In the employed Faster R-CNN method, the extracted feature maps from the FPN block are received, and anchors are generated with the anchor generator block. The generated anchors are in sizes 32, 64, 128, 256, and 512, with 0.5, 1.0, and 2.0 aspect ratios, which would also lead to providing a condition in which the model could be able to detect nodules of different sizes, especially small ones. In the next step, the Region Proposal Network (RPN) produces regions of interest (RoI) which are locations on the feature maps with a high probability of containing objects. These RoIs are then passed to the Multi-Scale Region of Interest Alignment (MS-RoIAlign) to determine the most appropriate feature maps from which RoIs should be used, followed by extracting the fixed-size spatial resolution of the RoIs. When a RoI is defined on a feature map, it typically doesn't align perfectly with the pixel grid, hence, the MS-RoIAlign uses bilinear interpolation to ensure that the RoIs' features are aligned regardless of the spatial resolution of the feature map from which they were extracted. The extracted RoIs are then passed to the classification and regression heads to fit the bounding boxes and predict the class. Some detection samples are shown in Fig. 3 with their ground truths.

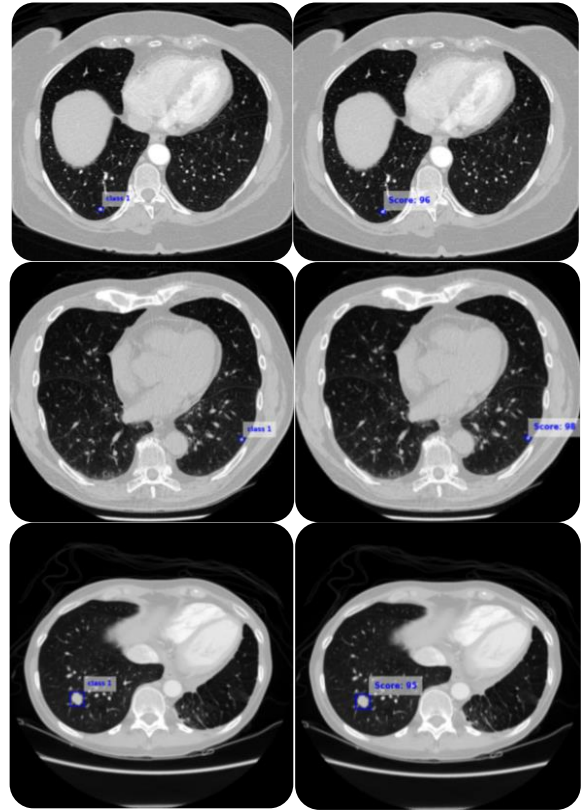

Fig 3. Left: ground truth of the nodules. Right: Our model's detection.

## IV. DATASET

The LUNA16 dataset is used for lung nodule detection in this paper. The LUNA16 dataset is an extensive resource and reference for the development and evaluation of automated lung nodule detection systems in CT images. This dataset was created to provide a standard framework for evaluating pulmonary nodule detection algorithms, and it consists of 888 CT scans with 1186 nodules, which are extracted from the LIDC-IDRI dataset. The slices have 512×512×1 dimensions and each scan has 100 to 500 axial slices.

## V. EXPERIMENTS AND RESULTS

### A. Implementation Settings

The experiments were conducted on Google Collaboratory. The model and training procedure were carried out mainly using the PyTorch library. Transfer Learning was considered as the training procedure, by which we fine-tuned the pre-trained models using the prepared LUNA16 dataset. The employed Swin-T model was an ImageNet pre-trained model provided by the *PyTorch* library. The total training epoch was only 5, which was significantly lower compared to the previous works, and the optimizer was Stochastic Gradient Descent with an initial learning rate of 0.005, momentum of 0.9, and weight decay of 0.0001. Moreover, a linear learning rate warmup was utilized with 1000 iterations and a 0.001 warmup factor. All the previous works that are presented for comparison were pre-trained models on ImageNet or COCO, which were fine-tuned with the prepared LUNA16 dataset under the same training settings. Also, all the results are averaged over 5 executions to provide robust results.

### B. Evaluation metrics

The proposed model is evaluated using average precision (AP) and average recall (AR) under different nodule area sizes, which are categorized into small, medium, large, and all. The nodules were sorted by area size and then categorized into four categories. Evaluation is on the Intersection Over Union (IoU) thresholds from 0.5 to 0.95 with 0.05 step size, which is 0.5, 0.55, 0.6, 0.65, 0.7, 0.75, 0.8, 0.85, 0.9, and 0.95, resulting in 10 AP and AR values in total. The total AP and AR values under every IoU threshold are then averaged, resulting in single mean average precision (mAP) and mean average recall (mAR). The calculation of mAP consists of calculating AP for every IoU threshold at different recall levels, and with (1), the area under the Precision-Recall (PR) curve is calculated.

$$AP_t = \int_0^1 P(R)dR \quad (1)$$

Subsequently, the mAP is computed by using (2) where the N represents the number of APs under all of the IoU thresholds.

$$mAP = \frac{1}{N} \sum_{t \in \{0.5,\ldots 0.95\}} AP_t \quad (2)$$

Average Recall is also computed by calculating the recall for every IoU threshold by using (3), then averaging over all the computed ARs with (4), where $M$ is the number of IoU thresholds.

$$R_{t,D} = \frac{True\ Positive_D}{True\ Positive_D + False\ Negative_D} \quad (3)$$

$$AR_D = \frac{1}{M} \sum_{t \in \{0.5,\ldots 0.95\}} R_{t,D} \quad (4)$$

### C. Comparison and Evaluation

The proposed model is compared to a baseline model and two state-of-the-art models. As shown in Table III, the performance gap between our model and two of the other works, [14] and [9] was significant, e.g. for small nodules, there was a 19.3% gap between our model and the model presented in [9]. The mAP and mAR of the proposed model under all nodule area sizes is slightly higher than the SOTA model in [16] with 0.2%. Also, it attained the highest mAP and mAR for small nodules compared to the same model by 1.3% and 1.6% respectively. However, the model in [16] performed better in detecting large nodules with 99.3% and slightly in medium-sized nodules with 96.6% of mAP, which are higher than our model by 6.5% and 0.1% respectively.

TABLE III. COMPARISON OF OUR MODEL WITH OTHER MODELS IN TERMS OF MAP AND MAR OF IOU THRESHOLD OF 0.5 TO 0.95 WITH 0.05 STEP SIZE UNDER DIFFERENT NODULE AREA SIZES

| Model | Area | mAP | mAR |
|---|---|---|---|
| **Ren et al. [14]** | Small | 0.704 | 0.706 |
| | Medium | 0.853 | 0.859 |
| | Large | 0.846 | 0.867 |
| | All | 0.797 | 0.803 |
| **Liu et al. [9]** | Small | 0.719 | 0.720 |
| | Medium | 0.859 | 0.861 |
| | Large | 0.901 | 0.906 |
| | All | 0.807 | 0.810 |
| **Li et al. [16]** | Small | 0.899 | 0.901 |
| | Medium | **0.966** | **0.974** |
| | Large | **0.993** | **0.994** |
| | All | 0.944 | 0.947 |
| **Ours** | Small | **0.912** | **0.917** |
| | Medium | 0.965 | 0.973 |
| | Large | 0.928 | 0.936 |
| | All | **0.947** | **0.949** |

TABLE IV. PERFORMANCE OF OUR MODEL IN DIFFERENT IOU THRESHOLDS AND NODULE AREA SIZES

| Area | IoU threshold | AP | AR |
|---|---|---|---|
| Small | 0.50 | 0.926 | 0.934 |
| | 0.75 | 0.921 | 0.928 |
| | 0.95 | 0.921 | 0.925 |
| Medium | 0.50 | 0.970 | 0.978 |
| | 0.75 | 0.970 | 0.978 |
| | 0.95 | 0.970 | 0.978 |
| Large | 0.50 | 0.969 | 0.972 |
| | 0.75 | 0.960 | 0.970 |
| | 0.95 | 0.960 | 0.968 |
| All | 0.50 | 0.960 | 0.960 |
| | 0.75 | 0.950 | 0.958 |
| | 0.95 | 0.950 | 0.958 |

As described, the proposed model achieved competitive results compared to the state-of-the-art (SOTA) models, and even better in certain conditions e.g. small nodules and all-size nodules. Moreover, Table IV demonstrates the performance of our proposed model in more detail, by reporting the AP and AR in three distinct IoU thresholds of 0.5, 0.75, and 0.95 under four nodule area sizes of small, medium, large, and all. The model achieved AP and AR over 92.1% regardless of the IoU threshold and nodule area, which demonstrates a robust performance in detecting lung nodules. Also, the model performed over 96% in detecting large nodules over all of the IoU thresholds and over 95% in all size nodules regardless of the IoU thresholds.

## VI. CONCLUSION

In this paper, we proposed a lung nodule detection model with a much more efficient approach compared to the other existing works. Also, we aimed not only for improvements in the performance of the model, especially in detecting small nodules, but we also considered an efficient approach in every aspect of our proposed model. We utilized the 2D slices of 3D CT scans of the LUNA16 dataset, resulting in using 2D models, which are significantly more efficient than 3D models. Also, we employed a Swin Transformer tiny to use the Vision Transformers benefits while not having high computational complexity, followed by a Feature Pyramid Network to detect nodules with various sizes better. Moreover, we used Transfer Learning to fine-tune the model more efficiently, which resulted in a significantly lower training load. Ultimately, our proposed model achieved competitive results with the SOTA models and even better performance in detecting small nodules.